\documentclass[%
 reprint,superscriptaddress,
 amsmath,amssymb,prb,
]{revtex4-2}

\usepackage{graphicx}
\usepackage{dcolumn}
\usepackage{bm}
\usepackage{color}
\usepackage{siunitx}
\usepackage[version=3]{mhchem} 
\usepackage{xr}
\usepackage[normalem]{ulem}
\def \CGO{\ce{CuGeO3}}

\def \Tsp{$T_{\mathrm{sp}}$}
\def \Tb{$T<T_{\mathrm{sp}}$}
\def \Th{$T>T_{\mathrm{sp}}$}
\def \sqw{$S(q,\omega)$}

\begin{document}
\preprint{APS/123-QED}

\title{From weakly interacting spinons to tightly bound triplons in the frustrated quantum spin-Peierls chain}
\thanks{This manuscript has been authored by UT-Battelle, LLC under Contract No. DE-AC05-00OR22725 with the U.S. Department of Energy.  The United States Government retains and the publisher, by accepting the article for publication, acknowledges that the United States Government retains a non-exclusive, paid-up, irrevocable, world-wide license to publish or reproduce the published form of this manuscript, or allow others to do so, for United States Government purposes.  The Department of Energy will provide public access to these results of federally sponsored research in accordance with the DOE Public Access Plan (http://energy.gov/downloads/doe-public-access-plan).}

\author{Pyeongjae Park}
\email{parkp@ornl.gov}
\affiliation{Materials Science \& Technology Division, Oak Ridge National Laboratory, Oak Ridge, TN 37831, USA}

\author{Bo Xiao}
\affiliation{Materials Science \& Technology Division, Oak Ridge National Laboratory, Oak Ridge, TN 37831, USA}
\affiliation{Quantum Science Center, Oak Ridge, TN 37831, USA}

\author{Karolina G\'ornicka}
\affiliation{Materials Science \& Technology Division, Oak Ridge National Laboratory, Oak Ridge, TN 37831, USA}
\affiliation{Faculty of Applied Physics and Mathematics and Advanced Materials Centre, Gda\'nsk University of Technology, ul. Narutowicza 11/12, 80-233 Gda\'nsk, Poland} 

\author{Andrew F. May}
\affiliation{Materials Science \& Technology Division, Oak Ridge National Laboratory, Oak Ridge, TN 37831, USA}

\author{Jiaqiang Yan}
\affiliation{Materials Science \& Technology Division, Oak Ridge National Laboratory, Oak Ridge, TN 37831, USA}

\author{Ryoichi Kajimoto}
\affiliation{Materials and Life Science Division, J-PARC Center, Japan Atomic Energy Agency, Tokai, Ibaraki 319-1195, Japan}

\author{Mitsutaka Nakamura}
\affiliation{Materials and Life Science Division, J-PARC Center, Japan Atomic Energy Agency, Tokai, Ibaraki 319-1195, Japan}

\author{Matthew B. Stone}
\affiliation{Neutron Scattering Division, Oak Ridge National Laboratory, Oak Ridge, TN 37831, USA}

\author{G\'abor B. Hal\'asz}
\email{halaszg@ornl.gov}
\affiliation{Materials Science \& Technology Division, Oak Ridge National Laboratory, Oak Ridge, TN 37831, USA}

\author{Andrew D. Christianson}
\email{christiansad@ornl.gov}
\affiliation{Materials Science \& Technology Division, Oak Ridge National Laboratory, Oak Ridge, TN 37831, USA}

\begin{abstract}

Fractionalized quasiparticles and their confinement into emergent bound states lie at the heart of modern quantum magnetism. While the evolution into magnonic bound states has been well characterized, experimental insight into the analogous transition to triplons remains limited. Here, using high-resolution neutron spectroscopy and state-of-the-art spin dynamics simulations, we uncover the transformation from weakly interacting spinons to tightly bound triplons in the spin-Peierls compound \CGO{}. Quantitative comparisons between the measured spectra and tensor network simulations reveal substantial next-nearest-neighbor frustration and weak external dimerization, placing the system deep within the spontaneously dimerized regime and near the exactly solvable Majumdar–Ghosh point. We further show an energy- and temperature-dependent evolution between two contrasting quasiparticle regimes: deconfined spinons with markedly suppressed interactions by frustration, and coherent triplonic bound states with no observable spinon degrees of freedom. Remarkably, triplon character persists into the two-particle regime, forming a structured two-triplon continuum with a spectral feature associated with a van Hove singularity at its lower boundary. These findings challenge the conventional view that robust triplons require strong external dimerization and demonstrate how the interplay between frustration and dimerization can reshape fractionalization and confinement.
\end{abstract}

\maketitle

\begin{figure*}[ht!]
\includegraphics[width=0.95\textwidth]{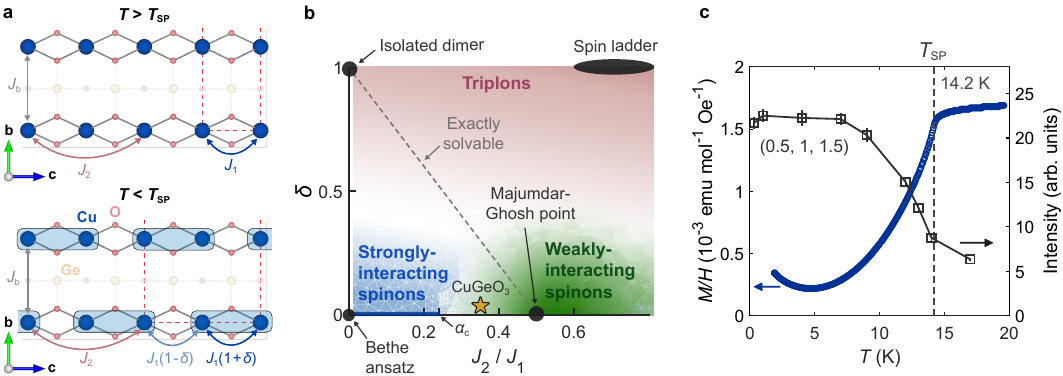} 
\caption{\label{basic} \textbf{Interplay between exchange frustration, dimerization, and excitation character in one-dimensional $\mathbf{S=1/2}$ AFM spin chains}. \textbf{a}, Crystal structure, minimal unit cells (red dashed lines), and corresponding interaction profiles of \CGO{} in the normal (\Th{}) and dimerized (\Tb{}) phases. \textbf{b}, Phase diagram of the $J_{1}-J_{2}-\delta$ spin model for $S=1/2$ chain systems, constructed based on prior theoretical studies~\cite{bethe1931, majumdar1969_1, shastry1981, haldane1982, okamoto1992, chitra1995}. The color code schematically illustrates the predominant types of elementary excitations in different regions, based on insights from three exactly solvable limits: the Bethe-ansatz limit ($J_{2}/J_{1}=\delta=0$)~\cite{bethe1931}, the isolated-dimer limit ($J_{2}/J_{1}=0$, $\delta=1$), and the Majumdar-Ghosh point ($J_{2}/J_{1}=0.5$, $\delta=0$)~\cite{majumdar1969_1, majumdar1969_2, shastry1981}. The thick blue horizontal line extending to $J_{2}/J_{1}=0.2411\equiv \alpha_{c}$ for $\delta=0$ denotes the Tomonaga-Luttinger liquid regime characterized by gapless deconfined spinons~\cite{haldane1981}. A star symbol denotes the location of \CGO{} determined in this study. \textbf{c}, Temperature dependence of the magnetic susceptibility and the intensity of the (0.5, 1, 1.5) superlattice peak associated with lattice distortion, demonstrating a spin-Peierls transition at \Tsp{}\,=\,14.2\,K.} 
\end{figure*}

Fractionalization---the breakdown of the original degrees of freedom into nonlocal quasiparticles carrying fractional quantum numbers---is a foundational principle of diverse quantum many-body phenomena~\cite{faddeev1981, laughlin1983, castelnovo2008}. A paradigmatic route to realizing fractionalized excitations is through antiferromagnetic (AFM) quantum spin ($S=1/2$) systems with magnetic frustration, which can host deconfined spinons~\cite{balents2010,savary2016}. In one dimension, deconfined spinons already emerge in the $S=1/2$ AFM nearest-neighbor interaction ($J_{1}$) model without frustration~\cite{bethe1931,balents2010}. Introducing AFM next-nearest-neighbor interactions ($J_2$) adds frustration to this strongly quantum system---effectively combining the two canonical mechanisms for spin fractionalization---and reveals a unique landscape of quantum phases and fractionalization phenomena. Haldane proposed that introducing AFM $J_2$ beyond a critical threshold $J_{2}/J_{1}>\alpha_c$ drives an exactly solvable quantum spin liquid (QSL) described by the Bethe ansatz~\cite{bethe1931} into a spontaneously dimerized ground state of spin singlets~\cite{haldane1982}. The critical threshold value $\alpha_{c}$ was later determined to be $0.2411$~\cite{okamoto1992}. The spin dimerization reduces translational symmetry and, via spin-lattice coupling, can induce a corresponding spin-Peierls transition that naturally leads to alternating $J_1$ couplings. The alternation can be parameterized by an external dimerization $\delta$, such that the intra- and inter-dimer interactions become $(1 \pm \delta)J_1$, respectively (see Fig.~\ref{basic}a).

The resulting $J_{1}-J_{2}-\delta$ model for $S=1/2$ AFM chains hosts a variety of elementary excitations, whose mutual transformations provide unique insights into the physics of fractionalization and confinement (see Fig.~\ref{basic}b). The microscopic spin Hamiltonian is expressed as: 

\begin{equation}
\hat{\cal H} = \sum_{i} 
\left\{J_{1}
  [\,1 + (-1)^{\,i+1}\delta]\;\hat{\mathbf{S}}_i \hat{\mathbf{S}}_{i+1}
  \;+\;J_{2}\;\hat{\mathbf{S}}_i \hat{\mathbf{S}}_{i+2}
\right\},
\label{eq:Hamiltonian}
\end{equation}
where $\hat{\mathbf{S}}_i$ is the spin-1/2 at site $i$. In the Bethe ansatz limit ($J_2/J_1 = \delta = 0$), excitations are strongly interacting deconfined spinons. In the strong-dimer limit ($\delta \approx1$), they evolve into $S=1$ triplons, which can be understood as two-spinon bound states---analogous to magnons in proximate-QSL systems~\cite{bera2017, ghioldi2018}. A third key regime is the Majumdar–Ghosh (M-G) limit ($J_2/J_1 = 0.5$, $\delta = 0$) featuring an exactly solvable valence bond solid ground state~\cite{majumdar1969_1, majumdar1969_2}, where Shastry and Sutherland showed within a simple variational framework that the excitations behave as weakly interacting spinons~\cite{shastry1981}. These limits together imply a transition between qualitatively distinct quasiparticle types---weakly and strongly interacting spinons, and triplons---potentially revealing a unique fractionalization and confinement process where nearly free spinons can dramatically pair into tightly bound triplons. However, despite extensive theoretical interest~\cite{caspers1984, brehmer1998, bouzerar1998, uhrig1996, uhrig1999, singh1999, byrnes1999}, detailed experimental tracking of such excitation characters and their spectral evolution has remained elusive.

\CGO{}, a quasi-one-dimensional (quasi-1D) $S=1/2$ AFM chain system of Cu$^{2+}$ ions (Fig.~\ref{basic}a), offers an ideal platform to investigate the intriguing physics outlined above. It undergoes a spin-Peierls transition at \Tsp{}\,=\,14.2\,K, stabilizing a dimerized singlet ground state with a finite excitation energy gap~\cite{hase1993, nishi1994, Pouget1994}, and has been extensively studied as a foundational model of quantum magnetism~\cite{nishi1994, Hirota1994, kuroe1994, castilla1995, riera1995, arai1996, regnault1996, khomskii1996, ain1997_DG, fabricius1998, bouzerar1999, fujita2013}. Magnetic susceptibility analyses suggested a sizable AFM $J_2$, placing \CGO{} as a rare example of frustrated $S=1/2$ chains~\cite{castilla1995, riera1995, fabricius1998, bouzerar1999}. However, estimated $J_2/J_1$ values vary widely from 0.2 to 0.36, straddling the critical threshold for spontaneous dimerization ($\alpha_{c} = 0.2411$). Thus, whether the spin-Peierls transition is driven purely by lattice effects or also significantly influenced by exchange frustration remains unresolved~\cite{buchner1996, fabricius1998}. Previous pioneering inelastic neutron scattering (INS) studies identified key spectral features of \CGO{}, including gapped dispersive modes and high-energy continua characteristic of dimerized $S=1/2$ AFM chains~\cite{nishi1994, arai1996, regnault1996, ain1997_DG, ikeuchi2013, fujita2013}. Analyzing the full energy- and momentum-dependent dynamical structure factor [$S(\mathbf{q},\omega)$] will enable accurate determination of interaction parameters and reveal the evolution between distinct excitation characters anticipated by the model (see Fig.~\ref{basic}b).

\begin{figure*}[ht]
\includegraphics[width=1\textwidth]{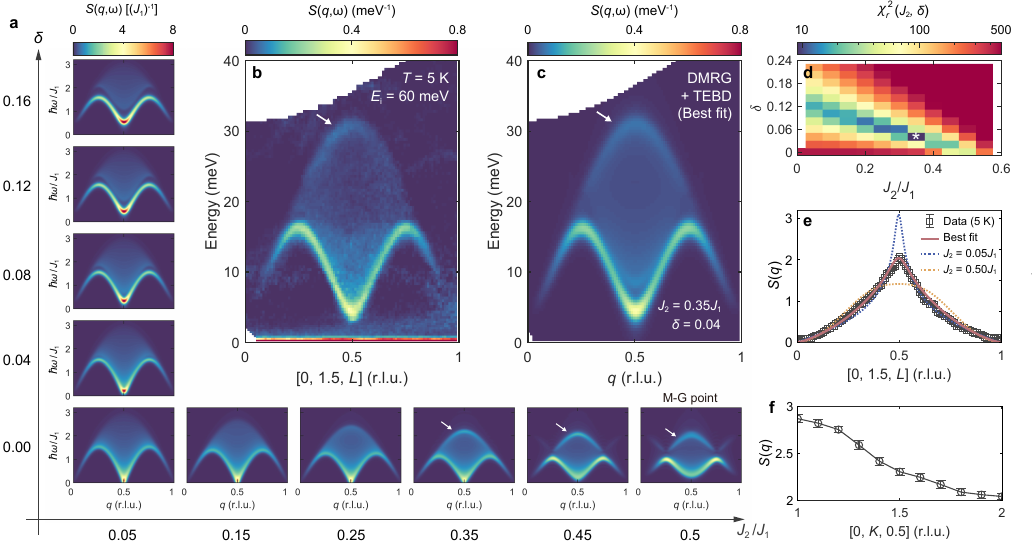} 
\caption{\label{DMRG} \textbf{Full excitation spectrum and corresponding spin Hamiltonian of \CGO{} at \Tb{}.} \textbf{a}, Dynamical structure factor [$S(q,\omega)$] maps of the $J_{1}-J_{2}-\delta$ spin model obtained from 1D DMRG\,+\,TEBD (see Methods) for various sets of $(J_{2}/J_{1}, \delta)$ parameters. \textbf{b}--\textbf{c}, $S(\mathbf{q},\omega)$ of \CGO{} at 5\,K (\Tb{}) and the corresponding best-fit DMRG\,+\,TEBD result, obtained from $J_{1} = 13.79(2)\,$meV, $J_{2}=0.35J_{1}$, and $\delta=0.04$. In \textbf{b}, high-resolution data from low incident neutron energy ($E_{i}=24$\,meV) is overlaid below 3.5\,meV to improve the presentation of the low-energy spectrum. White arrows in \textbf{a}--\textbf{c} indicate a coherent spectral structure along the upper edge of a two-spinon continuum. \textbf{d}, Reduced chi-square ($\chi_{r}^{2}$) comparing the measured and simulated \sqw{} across the 2D parameter space of $(J_{2}/J_{1}, \delta)$, shown on a logarithmic scale. The location of the minimum $\chi_{r}^{2}$ (= 9.195) is marked by an asterisk (*). \textbf{e}, Detailed comparison between the measured and calculated instantaneous structure factor [$S(q)$]. The solid line and the two dotted lines represent the calculated $S(q)$ for three different parameter sets: the optimal solution, $(J_{2}/J_{1}, \delta)=(0.05,0.04)$, and $(J_{2}/J_{1}, \delta)=(0.5,0.04)$. \textbf{f}, Instantaneous structure factor profile along the inter-chain [0, $K$, 0] direction at $L=0.5$. All structure factors are presented in absolute intensity units. Error bars represent standard deviations of the measured structure factors.}
\end{figure*}

In this work, we combine single-crystal INS with advanced spin dynamics simulations to refine the spin Hamiltonian of \CGO{} and uncover a spectral evolution from weakly interacting spinons to strikingly tightly bound triplons. We begin with benchmarking high-precision measurements of $S(\mathbf{q},\omega)$ at \Tb{} against state-of-the-art tensor network simulations, enabling precise extraction of $J_{2}/J_{1}$ and $\delta$. These values place \CGO{} deep within the spontaneously dimerized regime near the Majumdar-Ghosh point, whose physical realization has long remained elusive. With these parameters, we first describe the excitation character at high energies, which reveals a weakly interacting nature of deconfined spinons due to frustration. We then turn to high resolution measurements of the low-energy spectrum, which uncovers coherent, tightly bound triplons stabilized by weak external dimerization ($\delta\ll1$). Remarkably, in contrast to previous interpretations, the triplonic character persists into a two-particle continuum above the one-triplon branch, forming a structured continuum whose lower boundary follows the one-triplon dispersion and exhibits a spectral feature associated with a van Hove singularities (VHS). Finally, we describe the energy- and temperature-driven evolution between these contrasting quasiparticle regimes. Given $\delta\ll1$, our results challenge the conventional view that strong external dimerization ($\delta\sim1$) is necessary to stabilize tightly bound triplons. Together, these findings provide rare experimental insight into how frustration and dimerization jointly drive the unique confinement–deconfinement evolution of fractionalized quasiparticles.

\vspace{-8pt}
\subsection*{Determining the spin Hamiltonian}
\vspace{-10pt}
In \CGO{} above \Tsp{}, Cu$^{2+}$ ions form chains along the $c$-axis, with their primary interaction profile well captured by $J_{1}$ and $J_{2}$ (see Fig.~\ref{basic}a). Below \Tsp{}, the doubling of the unit cell along $c$ reduces the symmetry and allows for nonzero $\delta$. Although \CGO{} is nominally a 1D spin chain system, weak AFM interchain interactions ($J_b$) along the $b$-axis also play a non-negligible role in shaping its low-energy dynamics~\cite{cowley1996}. Our magnetic susceptibility measurements, shown together with the neutron diffraction intensity of the [0.5, 1, 1.5] reflection (see Fig.~\ref{basic}c), confirm a spin-Peierls transition at \Tsp{}\,=\,14.2\,K, consistent with previous reports~\cite{hase1993, nishi1994, Hirota1994, regnault1996}. 

\begin{figure*}[ht]
\includegraphics[width=1\textwidth]{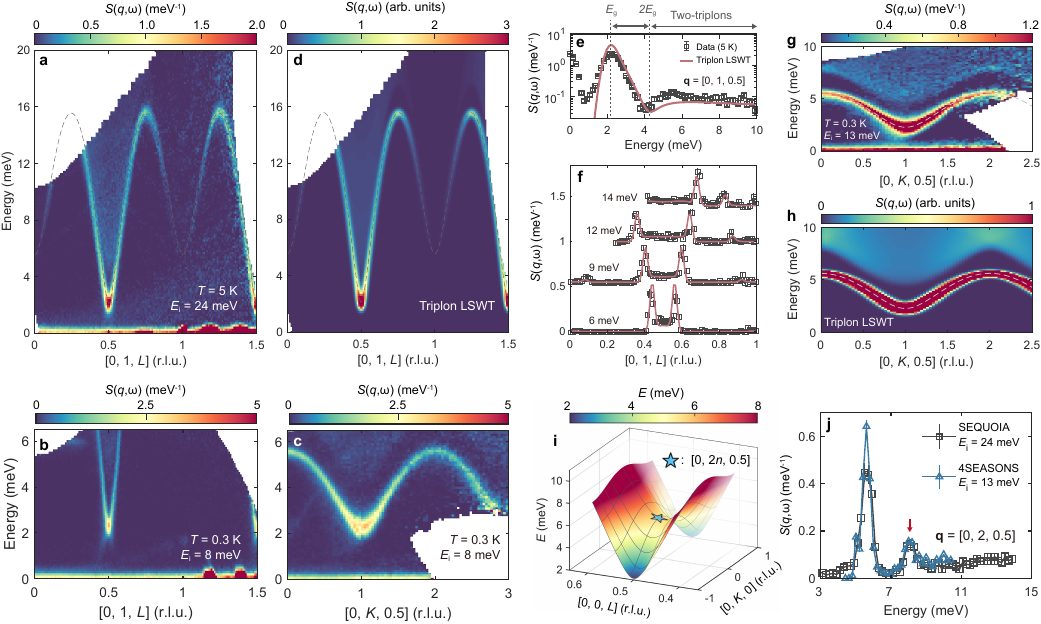} 
\caption{\label{LowE} \textbf{Highly coherent triplon modes in the low-energy excitation spectrum for \Tb{}.} \textbf{a}--\textbf{c}, High-resolution $S(\mathbf{q},\omega)$ spectra in the [0, \textit{K}, \textit{L}] plane, showing the gapped and dispersive excitations with ultrasharp energy linewidths. Data in \textbf{a} and \textbf{b}--\textbf{c} were collected at the SEQUOIA and 4SEASONS spectrometers, respectively. \textbf{d}, Theoretical $S(\mathbf{q},\omega)$ corresponding to \textbf{a}, calculated via triplon LSWT including instrumental resolution effects (see Methods). \textbf{e}--\textbf{f}, Intensity profiles as a function of energy or momentum transfer, comparing experiment and theory. \textbf{g}, Structured and gapped continuum excitations revealed by INS, uniformly separated from the one-triplon branch by its minimal energy gap of 2.1\,meV at $\mathbf{q}=(0,2n+1,0.5)$ ($n$ is an integer). \textbf{h}, Triplon LSWT result corresponding to \textbf{g}, with emphasis on the calculated two-triplon continuum. \textbf{i}, Two-dimensional dispersion of the one-triplon mode [$\omega(K,L)$], highlighting a van Hove singularity (VHS) at $\mathbf{q} = (0, 2n, 0.5)$. \textbf{j}, Energy-dependent intensities at $\mathbf{q}=[0,2,0.5]$, each measured at SEQUOIA (5\,K) and 4SEASONS (0.3\,K), highlighting a VHS-induced spectral peak (red arrow). Calculated $S(\mathbf{q},\omega)$ in \textbf{e}, \textbf{f}, \textbf{h}, and \textbf{j} are scaled by an arbitrary factor for comparison.}
\end{figure*}

To quantitatively determine $J_{2}/J_{1}$ and $\delta$ for $\mathrm{CuGeO_{3}}$, we compare the $S(q, \omega)$ spectra measured by INS with high-precision tensor network simulations (Fig.~\ref{DMRG}). The ground state of the $J_{1}-J_{2}-\delta$ model is represented as a matrix product state (MPS) and variationally optimized using the density-matrix renomalization group (DMRG) algorithm~\cite{White1992, White1993}. Time-dependent spin-spin correlation functions are then computed through real-time evolution using the time-evolving block decimation (TEBD) algorithm~\cite{Vidal2004, White2004, Daley2004}, from which $S(q, \omega)$ is obtained through a double Fourier transform. To span the relevant parameter space, we performed DMRG\,+\,TEBD simulations across a broad grid of $J_2/J_{1}$ and $\delta$ values (see Supplementary Note~\ref{supp-sec:DMRG_fit}). As shown in Fig.~\ref{DMRG}a (see also Fig.~\ref{supp-DMRG_full}), increasing $\delta$ primarily opens a spin gap, while increasing $J_2/J_1$ introduces two unique spectral features in addition to a spin gap induced by $J_2/J_1>\alpha_c$: (i) spectral weight shifts away from $q = 0.5$\,(r.l.u.) (see Fig.~\ref{DMRG}e), and (ii) enhanced intensity appears along the upper edge of the two-spinon continuum (white arrows in Fig.~\ref{DMRG}). Meanwhile, the measured $S(\mathbf{q},\omega)$ shows variation along the [0, $K$, 0] inter-chain direction at low energies due to weak $J_b>0$ (see Fig.~\ref{LowE}). To enable direct comparison with the 1D tensor network simulations, we extract experimental cuts near $K = 1.5$, where the dispersion is independent of interchain modulation and thus the quasi-1D character is best represented (see Supplementary Note~\ref{supp-sec:DMRG_fit}).

We identify a global minimum in the reduced chi-square metric ($\chi_{\mathrm{r,min}}^2 = 9.195$) between the measurement and the calculations for $J_1 = 13.79(2)$\,meV, $J_2 = 0.35J_1$, and $\delta = 0.04$ (Fig.~\ref{DMRG}d). Importantly, this comparison is performed in absolute intensity units (meV$^{-1}$) without arbitrary scaling (see Supplementary Note~\ref{supp-sec:DMRG_fit}). The solution is statistically robust, as nearby parameter sets yield $\chi_\mathrm{r}^2$ values far exceeding $\chi_{\mathrm{r,min}}^2 + 1$. The corresponding DMRG\,+\,TEBD results show excellent agreement with the data (Fig.~\ref{DMRG}b--c), successfully capturing the spin gap, momentum dependence of the instantaneous structure factor $S(q)$ (Fig.~\ref{DMRG}e), and the coherent spectral structure along the top of the two-spinon continuum (white arrows in Fig.~\ref{DMRG}).

Our optimized value of $J_{2}/J_{1} = 0.35$ resolves the longstanding uncertainty regarding the degree of magnetic frustration in \CGO{}. This value agrees with some earlier estimates from magnetic susceptibility~\cite{riera1995, fabricius1998}, although the same approach also resulted in different $J_{2}/J_{1}$ values, varying between 0.2 and 0.36~\cite{castilla1995, bouzerar1999}. Crucially, $J_{2}/J_{1} = 0.35$ places \CGO{} well above the theoretical threshold for spontaneous dimerization, $\alpha_c = 0.2411$~\cite{haldane1982, okamoto1992}. This identifies the system, to our knowledge, as the only known frustrated $S = 1/2$ spin chain featuring $J_{2}/J_{1}>\alpha_{c}$ and lying near the M-G point. This finding strongly suggests that thermodynamic considerations of the spin-Peierls transition in \CGO{} should include both lattice effects and exchange frustration~\cite{buchner1996, fabricius1998}. The spontaneous breaking of translational symmetry by spin dimerization can induce a lattice distortion via spin-lattice coupling, which in \CGO{} is known to manifest predominantly as an alternating Cu–O–Cu bond angle, naturally leading to a nonzero $\delta$ below \Tb{}~\cite{Braden1996, khomskii1996}.

\vspace{-8pt}
\subsection*{Excitation character at high energies}
\vspace{-10pt}
The high-energy spectrum of \CGO{} exhibits the characteristic two-spinon continuum of $S=1/2$ AFM chains (see Fig.~\ref{DMRG}b), indicating deconfined spinons as the elementary quasiparticles. At the same time, it also displays a coherent intensity enhancement along the continuum's upper edge (white arrows in Fig.~\ref{DMRG})~\cite{fujita2013}, a feature absent in $J_{1}$-only AFM spin chains. While the physical origin of this feature has previously remained elusive, our DMRG\,+\,TEBD analysis combined with the optimized Hamiltonian indicates that it arises from the suppressed interactions between spinons. As shown in Fig.~\ref{DMRG}a (also see Fig.~\ref{supp-DMRG_full}), this coherent structure becomes more pronounced with increasing $J_2/J_1>0$ and remains prominent around the M-G point, at which the spinons are weakly interacting~\cite{shastry1981}. This is distinct to the strongly interacting deconfined spinons found in the Bethe ansatz limit (see Fig.~\ref{basic}b), and therefore, the coherent feature at the upper edge can be interpreted as a hallmark of suppressed spinon–spinon interactions near the M-G regime.

This interpretation is also supported by a simple physical picture. In the non-interacting limit, the spinon contribution to $S(q,\omega)$ is directly proportional to the two-particle density of states (DOS), which exhibits a square-root divergence at the continuum’s upper edge for all momenta. Indeed, a similar coherent structure appears in the $S=1/2$ XY model that supports exactly non-interacting spinons~\cite{caux2011}. This consistency reinforces the conclusion that the high-energy excitations in \CGO{} are governed by deconfined but weakly interacting spinons. While they retain the general continuum profile of deconfined spinons expected for $S=1/2$ AFM chains, they qualitatively differ from the $J_1$-only case where strong spinon-spinon interactions suppress the coherent upper-edge structure associated with singular two-particle DOS~\cite{caux2011}.

\vspace{-8pt}
\subsection*{Excitation character at low energies}
\vspace{-10pt}
Figs.~\ref{LowE}a--c show the high-resolution $S(\mathbf{q}, \omega)$ spectra of \CGO{} below \Tsp{}, presented in absolute intensity units. A well-defined, dispersive excitation branch is observed both along the spin-chain direction (see Figs.~\ref{LowE}a–b) and perpendicular to it (see Fig.~\ref{LowE}c), indicating a 2D character of the low-energy excitations associated with both $J_{1,2}$ and $J_{b}$ (see Fig.~\ref{basic}a). While previous studies already reported a similar dispersion and interpreted it as a one-triplon branch~\cite{nishi1994, regnault1996, cowley1996, ain1997_DG}, our full intensity profiles also reveal the exceptionally narrow energy linewidths of the excitations. Theoretical $S(\mathbf{q},\omega)$ spectra calculated using triplon LSWT with instrumental resolution convolution, with interaction parameters chosen to reproduce the observed dispersion (see Methods and Supplementary Note~\ref{supp-sec:1triplon} and \ref{supp-sec:res}), yield comparable energy linewidths (see Fig.~\ref{LowE}d--f), strongly suggesting long lifetimes of the low-energy triplons in \CGO{}. Notably, as shown in prior studies~\cite{nishi1994, regnault1996, cowley1996} and discussed in Supplementary Note~\ref{supp-sec:1triplon}, LSWT fitting of the dispersion significantly underestimates $J_2/J_1$ compared to our DMRG\,+\,TEBD-based result of 0.35, due to its overestimation of the gap-opening effect by $J_2/J_1$. Nevertheless, despite not capturing the true spin Hamiltonian, triplon LSWT remains a useful semi-quantitative tool for modeling the low-energy spin dynamics in \CGO{}, particularly the resolution-limited broadness of long-lived triplons.

More intriguingly, the low-energy spectra exhibit a well-structured continuum, suggesting pronounced multi-particle dynamics. Figures~\ref{LowE}b--c and especially Fig.~\ref{LowE}g demonstrate additional continuum excitations, clearly separated from the one-triplon branch without any spectral overlap. Remarkably, the lower boundary of the continuum follows the same dispersion as the one-triplon branch, maintaining a constant energy gap across momentum space. The size of this gap matches the minimum spin gap of the one-triplon branch (2.1\,meV) at $\mathbf{q} = [0, 2n+1, 1/2]$\,(r.l.u.). Another striking feature is the appearance of enhanced spectral weight at the lower edge of the continuum near $\mathbf{q} = [0, 2n, 1/2]$\,(r.l.u.). This feature is further highlighted in Fig.~\ref{LowE}j by a clear peak around 8\,meV at $\mathbf{q}=[0, 2, 1/2]$. The consistent observation of this coherent continuum-edge feature in two independent INS experiments supports its intrinsic origin. 


\begin{figure*}[ht]
\includegraphics[width=0.94\textwidth]{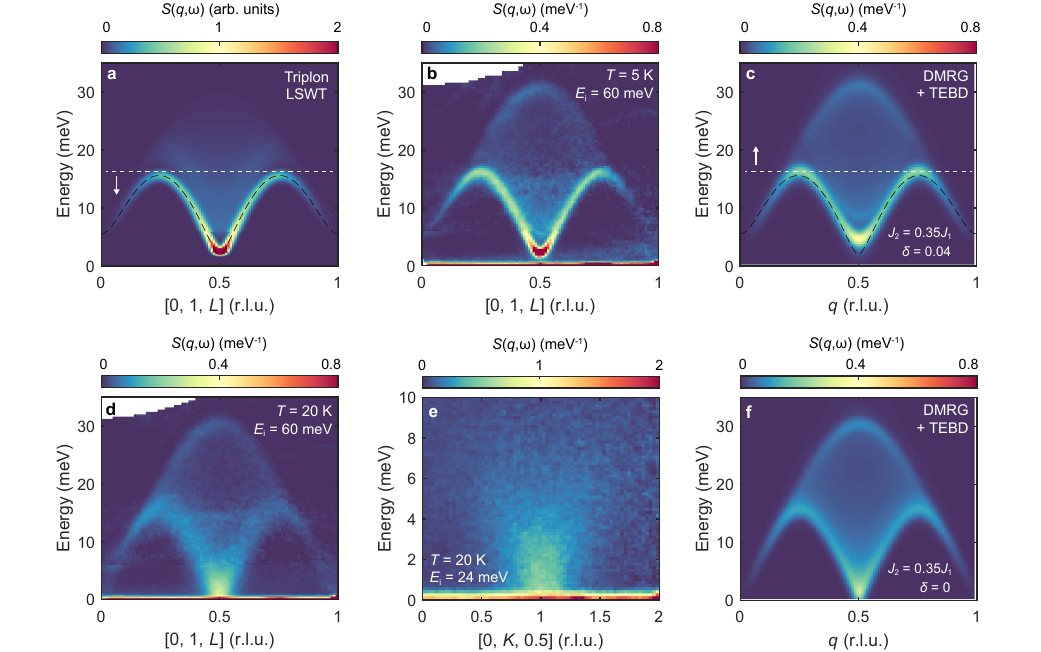} 
\caption{\label{crossover} \textbf{Confinement-deconfinement crossover of spinons across energy and temperature scales.} \textbf{a}--\textbf{c}, Comparison of measured $S(\mathbf{q},\omega)$ along the chain direction with theoretical predictions from triplon LSWT and DMRG\,+\,TEBD. Panel \textbf{c} shows the same result as Fig.~\ref{DMRG}c. Black dashed lines indicate the best-fit one-triplon dispersion obtained from triplon LSWT (see Supplementary Note~\ref{supp-sec:1triplon}). White horizontal dashed lines in \textbf{a} and \textbf{c} mark an empirical crossover energy, below (above) which the 2D triplon description (the 1D spinon description) better captures the data. \textbf{d}--\textbf{e}, $S(\mathbf{q},\omega)$ spectra measured above \Tsp{}. \textbf{f}, DMRG\,+\,TEBD spectrum with the same $J_{1}$ and $J_{2}$ as in \textbf{c}, but with $\delta=0$. In \textbf{b} and \textbf{d}, high-resolution data from low incident neutron energy ($E_{i}=24$\,meV) is overlaid below 6\,meV to improve the presentation of the low-energy spectrum. The calculations in this figure include instrumental resolution effects.}
\end{figure*}

It is noteworthy that a previous study, which lacked full momentum dependence of the continuum structure, attributed the continuum to deconfined spinons~\cite{ain1997_DG}. Yet the observed features in our data evidence its triplonic origin, highlighting the robustness of the triplon picture to fairly high energies. Primarily, the uniform energy gap between the one-triplon branch and the continuum is naturally explained by a two-triplon scenario, through simple kinematic constraints relating one- and two-triplon states in a non-interacting picture---even without detailed modeling~\cite{uhrig1996}. Indeed, our explicit triplon LSWT calculations including both one- and two-triplon states accurately reproduce the continuum structure (see Fig.~\ref{LowE}h). In contrast, interpreting the continuum as two-spinon excitations is problematic: spinon and triplon dispersions differ under the same Hamiltonian, making it unlikely for a two-spinon continuum to closely follow the one-triplon dispersion. 

Additional support for the two-triplon interpretation comes from the fact that the momentum $\mathbf{q} = [0, 2n, 1/2]$ of the coherent continuum-edge feature coincides with a VHS in both the one-triplon dispersion (see Fig.~\ref{LowE}i) and the joint two-triplon dispersion. This saddle-point structure translates into a logarithmic divergence near the lower edge of the two-triplon joint DOS, to which the continuum component of $S(\mathbf{q},\omega)$ is proportional in the non-interacting picture. Our triplon LSWT calculations again capture this effect by reproducing the enhanced spectral weight around $\mathbf{q} = [0, 2n, 1/2]$ near the onset of the continuum (see Fig.~\ref{LowE}h). This behavior is analogous to coherent structures observed in the two-magnon continua of magnetically ordered 2D systems~\cite{sala2021}, but represents the first experimental observation of a VHS in a two-triplon continuum.

The robustness of the triplon picture indicates that spinons are tightly bound into triplonic states at sufficiently low energies. This can be further substantiated by comparison between our results and the field-induced spinon-to-magnon confinement observed in an Ising spin chain~\cite{Woodland2023}. In the weak-confinement limit of Ref.~\cite{Woodland2023}, the two spinons comprising a magnonic bound state are separated by distances much larger than the lattice spacing. This gives rise to an entire series of sharp modes below the continuum that correspond to various internal excitations of the bounded two spinons. However, in the strong confinement limit, the two spinons cannot separate beyond one lattice spacing, rendering the individual spinons no longer observable, and yielding a spectrum fully understood in terms of one-magnon and two-magnon states alone~\cite{Woodland2023}.

The presence of a one-triplon mode and a kinematically matching two-triplon continuum in the low-energy spectrum of \CGO{} indicate that a pure triplon description is sufficient, without the need to account for the individual spinons inside the bound state. In other words, the low-energy spectrum of \CGO{} corresponds to the strong-confinement scenario above, with spinons tightly bound into triplons on the scale of a single lattice spacing. This finding is surprising given the weak external dimerization ($\delta = 0.04$), far from the $\delta \approx 1$ regime where the triplon description becomes exact (see Fig.~\ref{basic}b). In such a weakly dimerized regime, one would expect weak spinon confinement and corresponding multiple triplonic bound states reflecting the internal spinon degrees of freedom. Yet \CGO{} exhibits remarkably coherent triplon behavior, despite its weak external dimerization. 

\vspace{-7pt}
\subsection*{Energy- and temperature-dependent fractionalization}
\vspace{-8pt}

Figures~\ref{crossover}a--c summarize the energy-dependent evolution of the excitation character in \CGO{}. Below \Tsp{}, weak external dimerization ($\delta=0.04$) and AFM interchain coupling ($J_{b}$) freeze spin singlets into a static 2D array of staggered dimers, as shown in Fig.~\ref{basic}a. This ground state supports coherent 2D triplon modes and associated two-triplon continua---features captured by triplon LSWT. At higher energies, however, the excitation character shifts to deconfined spinons through fractionalization. Here, triplon LSWT fails to reproduce the coherent upper-edge structure in the two-spinon continuum, a key spectral feature of \CGO{}. Notably, this discrepancy between triplon LSWT and DMRG\,+\,TEBD in capturing the high-energy structure persists across a wide region of the $(J_{2}/J_{1},\delta)$ parameter space near the M-G limit (see Figs.~\ref{supp-DMRG_full} and \ref{supp-trip_full}), further supporting its spinon-based origin associated with their weak mutual interactions. We identify the top of the one-triplon branch (around 16\,meV) as an empirical crossover energy separating the two excitation characters (white dashed lines in Figs.~\ref{crossover}a and c). Notably, above this energy, the $S(\mathbf{q},\omega)$ spectrum shows no noticeable modulation along the [0, $K$, 0] inter-chain direction (see Fig.~\ref{supp-constE} and Supplementary Note~\ref{supp-sec:interchain}).

The temperature dependence of the spectrum across \Tsp{}\,=\,14.2\,K further illuminates the crossover behavior. Figures~\ref{crossover}d--e show $S(\mathbf{q},\omega)$ measured at 20\,K. The high-energy continuum remains largely unchanged, suggesting that spinons continue to be present and weakly interacting above \Tsp{}. At the same time, the low-energy triplon modes collapse into gapless, diffuse scattering---a change too abrupt to be explained by increased thermal fluctuations alone. Instead, we attribute this change to the loss of triplonic character, since translational symmetry is restored and $\delta$ vanishes immediately above \Tsp{} (see Fig.~\ref{basic}a). Supporting this interpretation, DMRG\,+\,TEBD calculations with the same $J_{1}$ and $J_{2}$ but $\delta = 0$ (see Fig.~\ref{crossover}f) provide a reasonable description of the 20\,K data. Here, the lower boundary of the two-spinon spectrum (the location of the coherent one-triplon branch for \Tb{}) merges into a continuum.

These results again suggest that even a small external dimerization ($\delta\ll1$) can act as a strong confinement potential, binding even weakly-interacting deconfined spinons into coherent $S=1$ triplons with no observable spinon degrees of freedom. Consequently, the triplonic character (reddish region in Fig.~\ref{basic}b) may dominate the low-energy excitations over a broader region of the $J_{2}/J_{1}-\delta$ phase diagram than naively conjectured from the three analytically solvable limits: isolated dimers, the Bethe ansatz, and the M-G point. This aspect enables \CGO{} to manifest a dramatic transformation between contrasting quasiparticle characters---nearly free spinons and tightly bound spinon pairs---within a single spin system, facilitated by low dimensionality and strong frustration.

The transformation pathway identified in \CGO{} raises broader questions about spinon-to-triplon confinement in dimerized 1D spin chains lacking significant frustration---i.e., spin-Peierls systems driven purely by lattice instabilities~\cite{bray1975, huizinga1979, cross1979}. In \CGO{}, where $J_2/J_1 > \alpha_c = 0.2411$, the ground state is already weakly dimerized (even without $\delta>0$), rather than forming a true Tomonaga-Luttinger liquid~\cite{haldane1981, haldane1982}. A key open question is whether introducing weak external dimerization ($\delta\ll1$) into a genuine Tomonaga-Luttinger liquid ($J_2/J_1 < \alpha_c$) could similarly stabilize tightly bound triplons and an associated two-triplon continuum, potentially revealing an even more dramatic change between liquid and solid valence-bond states. This question motivates further experimental efforts akin to the present study. High-resolution INS measurements on such systems are essential to map the full evolution landscape between spinon and triplon characters, which will ultimately complete our understanding of the excitation schemes across the entire $J_2/J_1–\delta$ phase space (see Fig.~\ref{basic}b). Overall, our results not only provide a comprehensive understanding of \CGO{}---a foundational model material in quantum magnetism---but also reveal, with unprecedented resolution, how the subtle interplay between frustration and dimerization reshapes fractionalized spinons and governs their confinement in low-dimensional quantum magnets.

\providecommand{\noopsort}[1]{}\providecommand{\singleletter}[1]{#1}%
\clearpage
\section*{Methods}

\subsection*{Sample preparation.}
High-quality \CGO{} single crystals were synthesized using a floating zone technique under carefully optimized conditions, based on the methods described in Refs.~\cite{revcolevschi1999, tanaka1996, dhalenne1997, watauchi2001}. The polycrystalline \CGO{} precursor was synthesized via a conventional solid-state reaction method using high-purity CuO and GeO$_{2}$ powders ($\geq99.99\,\%$) as starting materials, with an additional 5\,mol\,\% GeO$_{2}$ added to compensate for volatilization losses observed during growth. The mixed powders were initially calcined at 850$^{\circ}$C for 12 hours in air, followed by regrinding and a subsequent heat treatment at 850$^{\circ}$C for 24 hours. The resulting material was compacted into cylindrical feed rods ($\phi\approx5\,$mm, length $\approx 80\,$mm) via hydrostatic pressing and sintered at 950$^{\circ}$C for 48 hours in air. Single crystal growth was carried out in an infrared image furnace (NEC Nichiden Machinery Ltd., SC-M15HD) equipped with two halogen lamps as the radiation source. The growth was performed in an oxygen atmosphere at a rate of 2\,mm/h, with counter-rotation of the feed and seed rods at 30\,rpm to stabilize the molten zone. The initial as-grown crystals were used as seeds for growth along the [100] direction (normal to the natural cleavage plane). The resulting deep blue, semi-transparent crystals showed an elliptical cross section and were up to 100 mm long.

\subsection*{Bulk property measurements.}
Magnetization measurements were performed using a Quantum Design (QD) Magnetic Property Measurement System (MPMS3) SQUID magnetometer in vibrating sample magnetometry (VSM) mode. Measurements were made with the external field applied parallel to the $a$-axis. 

\subsection*{Single-crystal inelastic neutron scattering}
INS data were collected at the 4SEASONS time-of flight spectrometer at the Materials and Life Science Experimental Facility, J-PARC~\cite{4SEASONS2011}, and the SEQUOIA time-of-flight spectrometer at the Spallation Neutron Source, ORNL. For each experiment, three \CGO{} single crystals with total masses of 2.52\,g (4SEASONS) and 5.21\,g (SEQUOIA) were co-aligned on aluminum plates. The samples were mounted with the $(0KL)$ plane horizontal. Co-alignment was achieved with overall mosaicities of $2.5^{\circ}$ at 4SEASONS and $1.0^{\circ}$ at SEQUOIA (see Fig.~\ref{supp-coalign}). At 4SEASONS, data were collected using the repetition-rate-multiplication (RRM) technique~\cite{RRM2009}, which enabled simultaneous measurement with multiple incident neutron energies: $E_{\mathrm{i}} = 46.0$, 21.8, 12.6, and 8.3\,meV (chopper frequency: 200\,Hz). We used a \textsuperscript{3}He cryostat and acquired the data at 0.3, 1, 7, 17, 50, and 150\,K, with azimuthal sample rotation over a 130$^{\circ}$ range. The total counting time for each temperature is approximately 8 hours. To obtain a finer temperature-dependent data on the (0.5, 1, 1.5) superlattice peak intensity (see Fig.~\ref{basic}c), additional measurements were performed at 4, 9, 12, 13, and 14\,K using a 3$^{\circ}$ azimuthal sample rotation centered on the peak position. At SEQUOIA, we collected the data with $E_{\mathrm{i}}$ = 60 and 24\,meV under standard high-resolution chopper settings, at temperatures of 5, 20, 50, 100, and 150\,K, with a 120$^{\circ}$ range of sample rotation. The total counting time for each $E_{i}$ and temperature is approximately 6 hours, under the beam power of 1.7\,MW. 

We used the Horace~\cite{Horace} and Mantid~\cite{Mantid} software packages for analysis and visualization of four-dimensional $S(\mathbf{q},\omega)$ maps. All datasets were symmetrized according to the symmetry operations of the $Pnmm$ space group. Energy resolution profiles for each instrument and measurement condition are provided in Fig.~\ref{supp-Eres}. Momentum resolution of each dataset along the [0, $K$, 0] and [0, 0, $L$] was estimated from the full width at half-maximum (FWHM) of the [0, 1, 1] (r.l.u.) Bragg peak along each direction. These resolution profiles, along with the effects of finite integration width of momentum transfer perpendicular to the data cut direction (see Table~\ref{supp-tab:intrange} in Supplementary Note~\ref{supp-sec:interchain}), were used to convolve theoretical $S(\mathbf{q}, \omega)$ spectra obtained from DMRG\,+\,TEBD and triplon LSWT calculations. For SEQUOIA data, non-magnetic background--mostly phonon contributions--was estimated from high-momentum cuts ($L>1$) of the 150\,K dataset and subtracted. To ensure consistency with the 5\,K dataset, the 150\,K dataset were first corrected by dividing out the Bose factor to account for the detailed balance condition of $S(\mathbf{q},\omega)$ at finite temperature, before being subtracted as background.

The derivation of $S(\mathbf{q}, \omega)$ in absolute intensity units (meV$^{-1}$) from the measured INS cross-sections [which are not identical to $S(\mathbf{q}, \omega)$] is described in Supplementary Note~\ref{supp-sec:absolute}. The momentum integration ranges perpendicular to the plotting axis used for all $S(q,\omega)$ plots in the main text are summarized in Supplementary Table~\ref{supp-tab:intrange}.

\subsection*{Real-time evolution via time-evolving block-decimation}
To obtain the spectral functions, we evaluate time-dependent correlation functions by evolving quantum states along the real-time axis. The ground-state wavefunction of the dimerized $J_{1}-J_{2}$ Heisenberg chain is represented as a matrix product state (MPS) and variationally optimized using the density-matrix renomalization group (DMRG), enabling efficient truncation of the exponentially-large Hilbert space. Real-time dynamics are simulated using the time-evolving block decimation (TEBD) algorithm, which applies a sequence of local unitary gates to the perturbed MPS, enabling controlled and scalable time evolution of the system.

\vskip0.02in
The TEBD algorithm approximates real-time evolution by applying a Trotter-Suzuki decomposition to the unitary operator $U(t) = e^{-itH}$~\cite{Suzuki1976}, a method originally developed in the contexts of MPS~\cite{Vidal2004, Verstraete2004} and DMRG~\cite{White2004, Daley2004}. The time evolution is discretized into small time steps $\Delta\tau$, such that $U(\Delta\tau) = e^{i\Delta\tau H}$, allowing the Hamiltonian to be split into locally commuting terms and applied as efficient gate operations. While this introduces Trotter errors, they are systematically controlled by decreasing $\Delta\tau$ and employing higher-order decompositions. In this work, we employ a second-order Trotter scheme based on symmetrizing the decomposition of the original Hamiltonian. This yields a third-order error of $\mathcal{O}({\Delta\tau}^{3})$ per step and a second-order error of $\mathcal{O}({\Delta\tau}^{2})$ over the full evolution. 

\vskip0.02in
The ground-state wavefunction of the dimerized $J_{1}-J_{2}$ Heisenberg chain is computed using DMRG on a system of $N = 200$ sites, with parameters $(J_{2}/J_{1}, \delta/J_{1})$, and $J_{1} = 1$ set as the energy unit. A local excitation is introduced by applying the spin operator $S_{\ell}^{z}$ at site $\ell$, generating a localized wave packet composed of all wave vectors, which subsequently disperses under time evolution. The explicit dimerization breaks translational invariance, resulting in a two-site unit cell. As a consequence, spin-spin correlations are sensitive to the parity of the reference site. To capture this sublattice dependence, we independently time-evolve two perturbed states generated by applying $S_{\ell}^{z}$ at site $\ell = N/2 = 100$ and $\ell = N/2 + 1 = 101$, respectively. Each state is evolved up to a final time $T = 100$ (in units of $J_{1}$). To construct the time-evolution operator, the Hamiltonian is split into internally-commuting parts, each of which is exponentiated to from matrix product operators (MPOs). Time evolution proceeds by successively applying these MPOs to the MPS. Time-dependent spin-spin correlation functions are evaluated as
\begin{eqnarray}
    \langle S^{z}_{j}(t) S^{z}_{\ell}(0) \rangle = \langle 0 | U^{\dag}(t)S^{z}_{j} U(t) S^{z}_{\ell} | 0 \rangle,
\end{eqnarray}
where $|0\rangle$ denotes the ground state. The partial dynamical spin structure factor $S^{\mathrm{even}}(q, \omega)$, computed with the even central site as the reference, is obtained via a double Fourier transform of the time-dependent spin-spin correlation functions:
\begin{widetext}
    \begin{eqnarray}
        \begin{aligned}
            S^{\mathrm{even}}(q, \omega) &= \frac{1}{N} \sum_{j = 1}^{N} e^{-iq(j-N/2)} \int_{-\infty}^{\infty} dt \; e^{i\omega t} \langle S_{j}^{z}(t)S_{N/2}^{z}(0) \rangle \\
            &\cong \frac{2\pi}{N T} \Delta\tau \sum_{j = 1}^{N} e^{-iq(j - N/2)} \sum_{p = 0}^{L} e^{i(\omega + i\eta)t_{p}} 2{\rm Re} \langle S_{j}^{z}(t_{p}) S_{N/2}^{z}(0) \rangle,
        \end{aligned}
    \end{eqnarray}
where the time coordinate is discretized as $t_{p} = p\Delta\tau$, and the integral is truncated at a finite, maximum evolution time $T = L\Delta\tau$. To suppress spectral leakage arising from the finite time window, an exponential damping factor $\eta > 0$ is introduced. Repeating this procedure for reference site $\ell = N/2 + 1 = 101$ yields an additional contribution $S^{\mathrm{odd}}$ to the total dynamical spin structure factor,
    \begin{eqnarray}
    S^{\mathrm{odd}}(q, \omega) \cong \frac{2\pi}{N T} \Delta\tau \sum_{j = 1}^{N} e^{-iq(j - N/2 - 1)} \sum_{p = 0}^{L} e^{i(\omega + i\eta)t_{p}} 2{\rm Re}\langle S_{j}^{z}(t_{p}) S_{N/2 + 1}^{z}(0) \rangle.
\end{eqnarray}
The complete dynamical structure factor is obtained by summing the two contributions from both even and odd reference sites,
\begin{eqnarray}
    S(q, \omega) = S^{\mathrm{odd}}(q, \omega) + S^{\mathrm{even}}(q, \omega).
\end{eqnarray}

\subsection*{Triplon linear spin-wave theory calculations}

We consider an array of spin chains with staggered dimerization [see Fig.~\ref{basic}(a)] and the Hamiltonian in Eq.~(\ref{eq:Hamiltonian}) with additional interchain interactions of strength $J_b$. We introduce three bosons---corresponding to three flavors of triplons---for each dimer through a generalized $\mathrm{SU}(4)$ spin-wave theory~\cite{Muniz2014},
\begin{align}
S_{\mathbf{r}_{\mathrm{L}}}^z &= \frac{1}{2} \left[ z_{\mathbf{r}}^{\dag} \left( 2S - x_{\mathbf{r}}^{\dag} x_{\mathbf{r}}^{\phantom{\dag}} - y_{\mathbf{r}}^{\dag} y_{\mathbf{r}}^{\phantom{\dag}} - z_{\mathbf{r}}^{\dag} z_{\mathbf{r}}^{\phantom{\dag}} \right)^{1/2} + \left( 2S - x_{\mathbf{r}}^{\dag} x_{\mathbf{r}}^{\phantom{\dag}} - y_{\mathbf{r}}^{\dag} y_{\mathbf{r}}^{\phantom{\dag}} - z_{\mathbf{r}}^{\dag} z_{\mathbf{r}}^{\phantom{\dag}} \right)^{1/2} z_{\mathbf{r}}^{\phantom{\dag}} + i \left( x_{\mathbf{r}}^{\dag} y_{\mathbf{r}}^{\phantom{\dag}} - y_{\mathbf{r}}^{\dag} x_{\mathbf{r}}^{\phantom{\dag}} \right) \right] \nonumber \\
&= \frac{1}{2} \sqrt{2S} \left( z_{\mathbf{r}}^{\phantom{\dag}} + z_{\mathbf{r}}^{\dag} \right) + \frac{i}{2} \left( x_{\mathbf{r}}^{\dag} y_{\mathbf{r}}^{\phantom{\dag}} - y_{\mathbf{r}}^{\dag} x_{\mathbf{r}}^{\phantom{\dag}} \right) + O \left( S^{-1/2} \right), \nonumber \\
S_{\mathbf{r}_{\mathrm{R}}}^z &= \frac{1}{2} \left[ -z_{\mathbf{r}}^{\dag} \left( 2S - x_{\mathbf{r}}^{\dag} x_{\mathbf{r}}^{\phantom{\dag}} - y_{\mathbf{r}}^{\dag} y_{\mathbf{r}}^{\phantom{\dag}} - z_{\mathbf{r}}^{\dag} z_{\mathbf{r}}^{\phantom{\dag}} \right)^{1/2} - \left( 2S - x_{\mathbf{r}}^{\dag} x_{\mathbf{r}}^{\phantom{\dag}} - y_{\mathbf{r}}^{\dag} y_{\mathbf{r}}^{\phantom{\dag}} - z_{\mathbf{r}}^{\dag} z_{\mathbf{r}}^{\phantom{\dag}} \right)^{1/2} z_{\mathbf{r}}^{\phantom{\dag}} + i \left( x_{\mathbf{r}}^{\dag} y_{\mathbf{r}}^{\phantom{\dag}} - y_{\mathbf{r}}^{\dag} x_{\mathbf{r}}^{\phantom{\dag}} \right) \right] \nonumber \\
&= -\frac{1}{2} \sqrt{2S} \left( z_{\mathbf{r}}^{\phantom{\dag}} + z_{\mathbf{r}}^{\dag} \right) + \frac{i}{2} \left( x_{\mathbf{r}}^{\dag} y_{\mathbf{r}}^{\phantom{\dag}} - y_{\mathbf{r}}^{\dag} x_{\mathbf{r}}^{\phantom{\dag}} \right) + O \left( S^{-1/2} \right), \label{eq-varepsilon}
\end{align}
where $S=1/2$ is the spin length, while $\mathbf{r}_{\mathrm{L}}$ and $\mathbf{r}_{\mathrm{R}}$ are the left and right sites of the dimer centered at $\mathbf{r}$. Note that the $x$ and $y$ components are obtained by cyclic permutations of $x,y,z$. Introducing momentum-space triplon operators,
\begin{equation}
x_{\mathbf{r}} = \frac{1}{\sqrt{N}} \sum_{\mathbf{k}} e^{i \mathbf{k} \cdot \mathbf{r}} \, \tilde{x}_{\mathbf{k}}, \qquad y_{\mathbf{r}} = \frac{1}{\sqrt{N}} \sum_{\mathbf{k}} e^{i \mathbf{k} \cdot \mathbf{r}} \, \tilde{y}_{\mathbf{k}}, \qquad z_{\mathbf{r}} = \frac{1}{\sqrt{N}} \sum_{\mathbf{k}} e^{i \mathbf{k} \cdot \mathbf{r}} \, \tilde{z}_{\mathbf{k}},
\label{eq-z-1}
\end{equation}
the spin Hamiltonian can then---modulo an irrelevant constant term---be written as
\begin{equation}
H = 2S H^{(2)} + (2S)^{1/2} H^{(3)} + O(S^0), \label{eq-H-1}
\end{equation}
where the two-boson and three-boson terms are given by
\begin{align}
H^{(2)} &= J_0 \sum_{\mathbf{k}} \tilde{z}_{\mathbf{k}}^{\dag} \tilde{z}_{\mathbf{k}}^{\phantom{\dag}} - \frac{1}{4} \sum_{\mathbf{k}} \hat{J}_{\mathbf{k}} \left( \tilde{z}_{\mathbf{k}}^{\dag} + \tilde{z}_{-\mathbf{k}}^{\phantom{\dag}} \right) \left( \tilde{z}_{\mathbf{k}}^{\phantom{\dag}} + \tilde{z}_{-\mathbf{k}}^{\dag} \right) + [\textrm{cyclic permutations of } x,y,z], \nonumber \\
H^{(3)} &= \frac{i}{2\sqrt{N}} \sum_{\mathbf{k}, \mathbf{p}} \bar{J}_{\mathbf{k}} \left( \tilde{z}_{\mathbf{k}}^{\dag} + \tilde{z}_{-\mathbf{k}}^{\phantom{\dag}} \right) \left( \tilde{x}_{-\mathbf{p}}^{\dag} \, \tilde{y}_{\mathbf{k} - \mathbf{p}}^{\phantom{\dag}} - \tilde{y}_{\mathbf{p} - \mathbf{k}}^{\dag} \tilde{x}_{\mathbf{p}}^{\phantom{\dag}} \right) + [\textrm{cyclic permutations of } x,y,z] \label{eq-H-2}
\end{align}
in terms of the shorthand notations
\begin{align}
& J_0 = J_1 (1 + \delta), \quad \,\,\, \hat{J}_{\mathbf{k}} = \left[ J_1 (1 - \delta) - 2 J_2 \right] \cos (4 \pi L_{\mathbf{k}}) + 2 J_b \cos (\pi K_{\mathbf{k}}) \cos (2 \pi L_{\mathbf{k}}), \nonumber \\
& \bar{J}_{\mathbf{k}} = J_1 (1 - \delta) \sin (4 \pi L_{\mathbf{k}}) + 2 J_b \cos (\pi K_{\mathbf{k}}) \sin (2 \pi L_{\mathbf{k}}) \label{eq-J}
\end{align}
with standard momentum coordinates $\mathbf{k} = (H_{\mathbf{k}}, K_{\mathbf{k}}, L_{\mathbf{k}})$. Next, the quadratic two-boson term $H^{(2)}$ can be diagonalized by standard Bogoliubov transformations,
\begin{align}
& \tilde{x}_{\mathbf{k}}^{\phantom{\dag}} = u_{\mathbf{k}} X_{\mathbf{k}}^{\phantom{\dag}} + v_{\mathbf{k}} X_{-\mathbf{k}}^{\dag}, \qquad \tilde{y}_{\mathbf{k}}^{\phantom{\dag}} = u_{\mathbf{k}} Y_{\mathbf{k}}^{\phantom{\dag}} + v_{\mathbf{k}} Y_{-\mathbf{k}}^{\dag}, \qquad \tilde{z}_{\mathbf{k}}^{\phantom{\dag}} = u_{\mathbf{k}} Z_{\mathbf{k}}^{\phantom{\dag}} + v_{\mathbf{k}} Z_{-\mathbf{k}}^{\dag}, \nonumber \\
& u_{\mathbf{k}} = \cosh \phi_{\mathbf{k}}, \qquad v_{\mathbf{k}} = \sinh \phi_{\mathbf{k}}, \qquad \tanh 2 \phi_{\mathbf{k}} = \hat{J}_{\mathbf{k}} \left( 2 J_0 - \hat{J}_{\mathbf{k}} \right)^{-1},
\label{eq-z-1}
\end{align}
and the resulting triplon excitations $X_{\mathbf{k}}$, $Y_{\mathbf{k}}$, and $Z_{\mathbf{k}}$ have energy dispersions $E_{\mathbf{k}} = 2S \varepsilon_{\mathbf{k}}$ in terms of
\begin{equation}
\varepsilon_{\mathbf{k}} = \sqrt{J_0 \left( J_0 - \hat{J}_{\mathbf{k}} \right)}. \label{eq-varepsilon}
\end{equation}
The two lowest-order components of the dynamical spin structure factor---from the perspective of a standard $1/S$ expansion---are the single-triplon mode, $S^{(1)} (\mathbf{q}, \omega)$, and the two-triplon continuum, $S^{(2)} (\mathbf{q}, \omega)$. We calculate each of these components to its lowest-order contribution in $1/S$, which is $O(S)$ for the single-triplon mode and $O(S^0)$ for the two-triplon continuum. Taking $S = 1/2$, the intensity of the single-triplon mode becomes
\begin{equation}
S^{(1)} (\mathbf{q}, \omega) = \frac{1}{4} \left( u_{\mathbf{q}} + v_{\mathbf{q}} \right)^2 \left[ 1 - \cos (2 \pi L_{\mathbf{q}}) \right] \delta \left( \omega - \varepsilon_{\mathbf{q}} \right), \label{eq-S-1}
\end{equation}
while that of the two-triplon continuum takes the form
\begin{equation}
S^{(2)} (\mathbf{q}, \omega) = \frac{1}{4N} \sum_{\mathbf{k}} \left\{ A_{\mathbf{q}, \mathbf{k}}^2 \left[ 1 + \cos (2 \pi L_{\mathbf{q}}) \right] + B_{\mathbf{q}, \mathbf{k}}^2 \left[ 1 - \cos (2 \pi L_{\mathbf{q}}) \right] - 2 A_{\mathbf{q}, \mathbf{k}} B_{\mathbf{q}, \mathbf{k}} \sin (2 \pi L_{\mathbf{q}}) \right\} \delta \left( \omega - \varepsilon_{\mathbf{k}} - \varepsilon_{\mathbf{q} - \mathbf{k}} \right) \label{eq-S-2}
\end{equation}
in terms of the shorthand coefficients
\begin{align}
& A_{\mathbf{q}, \mathbf{k}} = u_{\mathbf{k}} v_{\mathbf{q} - \mathbf{k}} - v_{\mathbf{k}} u_{\mathbf{q} - \mathbf{k}}, \nonumber \\
& B_{\mathbf{q}, \mathbf{k}} = \frac{1}{2} \left( u_{\mathbf{q}} + v_{\mathbf{q}} \right) \Big\{ \left( \varepsilon_{\mathbf{k}} + \varepsilon_{\mathbf{q} - \mathbf{k}} + \varepsilon_{\mathbf{q}} \right)^{-1} \Big[ \bar{J}_{\mathbf{q}} \left( u_{\mathbf{q}} + v_{\mathbf{q}} \right) \left( u_{\mathbf{k}} v_{\mathbf{q} - \mathbf{k}} - v_{\mathbf{k}} u_{\mathbf{q} - \mathbf{k}} \right) - \bar{J}_{\mathbf{k}} \left( u_{\mathbf{k}} + v_{\mathbf{k}} \right) \left( u_{\mathbf{q} - \mathbf{k}} v_{\mathbf{q}} - v_{\mathbf{q} - \mathbf{k}} u_{\mathbf{q}} \right) \nonumber \\
& \qquad \quad \,\,\, -\bar{J}_{\mathbf{q} - \mathbf{k}} \left( u_{\mathbf{q} - \mathbf{k}} + v_{\mathbf{q} - \mathbf{k}} \right) \left( u_{\mathbf{q}} v_{\mathbf{k}} - v_{\mathbf{q}} u_{\mathbf{k}} \right) \Big] - \left( \varepsilon_{\mathbf{k}} + \varepsilon_{\mathbf{q} - \mathbf{k}} - \varepsilon_{\mathbf{q}} \right)^{-1} \Big[ \bar{J}_{\mathbf{q}} \left( u_{\mathbf{q}} + v_{\mathbf{q}} \right) \left( u_{\mathbf{k}} v_{\mathbf{q} - \mathbf{k}} - v_{\mathbf{k}} u_{\mathbf{q} - \mathbf{k}} \right) \nonumber \\
& \qquad \quad \,\,\, - \bar{J}_{\mathbf{k}} \left( u_{\mathbf{k}} + v_{\mathbf{k}} \right) \left( u_{\mathbf{q} - \mathbf{k}} u_{\mathbf{q}} - v_{\mathbf{q} - \mathbf{k}} v_{\mathbf{q}} \right) - \bar{J}_{\mathbf{q} - \mathbf{k}} \left( u_{\mathbf{q} - \mathbf{k}} + v_{\mathbf{q} - \mathbf{k}} \right) \left( v_{\mathbf{q}} v_{\mathbf{k}} - u_{\mathbf{q}} u_{\mathbf{k}} \right) \Big] \Big\}. \label{eq-ABC}
\end{align}
Note that, since the two-triplon continuum appears at a higher order of $1/S$, the calculation of $S^{(2)} (\mathbf{q}, \omega)$ requires perturbative $1/S$ corrections in both the spin operators and the ground state.
\end{widetext}

\vspace{1cm}

\section*{Data Availability}
Raw data are available from the corresponding author upon request. 
\vspace{1cm}

\section*{Code Availability}
Custom codes used in this article are available from the corresponding authors upon request. 
\vspace{1cm}

\begin{acknowledgments}
We acknowledge Cristian D. Batista for helpful discussions. K.A.G. and A.F.M. thank A. Revcolevschi for valuable insights regarding crystal growth. This work was supported by the U.S. Department of Energy, Office of Science, Basic Energy Sciences, Materials Sciences and Engineering Division. This research used resources at the Spallation Neutron Source, a DOE Office of Science User Facility operated by the Oak Ridge National Laboratory. The beam time was allocated to SEQUOIA on proposal number IPTS-34783.1. The neutron scattering experiment at the Japan Proton Accelerator Research Complex (J-PARC) was performed under the user program (Proposal No. 2024A0170). The work of B.X. was supported by the U.S. Department of Energy, Office of Science, National Quantum Information Science Research Centers, Quantum Science Center. This research used resources of the National Energy Research Scientific Computing Center (NERSC), a DOE Office of Science User Facility supported by the Office of Science of the U.S. Department of Energy under Contract No. DE-AC02-05CH11231 using NERSC award ASCR-ERCAP0032461.
\end{acknowledgments}

\section*{Author contributions}
P.P., G.B.H., and A.D.C. conceived the project. K.G., A.F.M., and J.Y. synthesized the samples and measured the bulk properties. P.P., R.K., M.N., M.B.S., and A.D.C. conducted the inelastic neutron scattering experiments. P.P. analyzed the neutron scattering data. B.X. desinged and performed the tensor network simulations. G.B.H. conducted triplon LSWT calculations. All authors participated in the data interpretation and discussion. P.P., G.B.H., and A.D.C. wrote the manuscript with contributions from all authors.

\section*{Competing Interests}
The authors declare no competing interests. 

\vspace{1cm}

\noindent\textbf{Supplementary Information} is available for this paper at [website url].


\clearpage
\end{document}